\documentclass[useAMS,usenatbib,onecolumn]{mn2e}

\usepackage{mathtext,amssymb}
\usepackage{epsfig}
\usepackage{graphics}

\renewcommand{\vector}[1]{\ensuremath{\mathbf{#1}}}

\newcommand{\mdot}{\ensuremath{\dot{m}}}

\newcommand{\Msunyr}{\ensuremath{\mathrm M_\odot\, \mathrm yr^{-1}}}

\title[BH spin-down by truncated disc emission]{Black hole spin-down by truncated disc emission}
\author[Abolmasov]{P. Abolmasov\thanks{E-mail: pavel.abolmasov@gmail.com}\\
Sternberg Astronomical Institute, Moscow State University,
  Moscow, Russia 119992\\}
\begin{document}

\date{Accepted ---. Received ---; in
  original form --- }

\pagerange{\pageref{firstpage}--\pageref{lastpage}} \pubyear{2012}

\maketitle

\label{firstpage}

\begin{abstract}
The influence of disc radiation capture upon black hole rotational evolution
is negligible for radiatively inefficient discs. For the standard thin disc
model it is a slight but potentially important effect leading to the equilibrium
spin parameter value of $a_{eq}\simeq 0.998$. 
For optically thin discs, the fraction of
disc radiation captured by the black hole is however about two times
larger. In some disc radiation models, inner parts of the accretion flow are
optically thin, advection-dominated flows, and the thin disc ends at some
transition radius $R_{tr}$. The thermal energy of the disc
stored in trapped radiation is released at this radius. Angular distribution
of the radiation released at this radial photosphere facilitates its
capture by the black hole. For accretion rates close to critical and disc
truncation radius $R_{tr}\simeq (2\div 4) GM/c^2 $, radiation capture is most
efficient in spinning the black hole down that may lead to $a_{eq} \sim
0.996\div 0.997$ 
or less depending on the mass accretion rate. For an accretion flow
radiating some constant fraction $\epsilon$ of dissipated energy, the
equilibrium Kerr parameter is shown to obey the relation $1-a_{eq} \propto
\epsilon^{3/2}$ as long as $1-a_{eq} \ll 1$.  { Deviations from Keplerian
  law near the last stable orbit dominate over the radiation capture effect if
  they exceed $1\div 2\%$.}
\end{abstract}


\section{Introduction}

Black hole is probably the simplest astrophysical system described,
apart from its position and velocity, by only one
scalar (mass) and one vector parameter (angular momentum). Angular momentum
$\vector{J}$ of a black hole is convenient to normalize by its highest
possible value:

$$
J = \frac{GM^2}{c} a
$$

Here, $-1<a<1$ is dimensionless Kerr parameter that we will consider as a
positive scalar assuming that the accretion disc lies within the black hole
equatorial plane and all the matter being accreted 
has angular momentum collinear with that of the black hole. 
The primary reason for this is Bardeen--Petterson
effect \citep{BP75} that aligns the inner parts of an accretion flow with
the black hole spin. A tilted disc exchanges angular momentum with the black
hole through Lense--Thirring precession but it does not affect the absolute value
of the spin. 

In the case of equatorial disc, evolution of a
black hole is described by the two first-order equations for mass and for Kerr
parameter (see \citet{bardeen70}). As long as the mass and angular momentum
accreted by the black hole depend linearly on the mass accretion rate, the
evolution may be expressed in terms of spin parameter change with mass:

$$
\frac{da}{d\ln M} = \frac{c^3}{GM} \frac{L^\dagger}{E^\dagger}-2a
$$

Here, $L^\dagger$ and $E^\dagger$ are the net (per unit mass) angular momentum 
and energy of the matter absorbed by the hole. 
For the case of thin disc accretion of ideal matter
% with zero non-diagonal
%stress terms in its energy-stress tensor
% ::referee!
{ with no additional stress terms,} black hole spin
evolution proceeds towards the maximal possible value of $a=1$ and formally
even further. For black holes close to the extreme Kerr case, unexpected
effects such as radiation capture may play the main role in stopping the
spin-up. In particular, black hole should absorb stellar light and cosmic
microwave background. Net angular momentum of distant photons is around zero
(see section \ref{sec:asympta}) hence they simply dilute the rotational energy
of the accreting black hole by irreducible mass. { If accretion is present,
its impact and impact of its radiation upon rotation of the black hole are
evidently much stronger.}
%% ::referee:
% The equilibrium rotation
%parameter is very close to unity in this case ($1-a \lesssim 10^{-18}$).
% If we introduce $\delta a =
%1-a$, it will be of the order $\delta a \sim 10^{-18}$ for a supermassive 
%black hole surrounded by a nuclear star cluster and $\sim 10^{-33\div-30}$ or
%less for a supermassive black hole spun down by the CMB only. 
Since Kerr parameter is likely to differ from zero by a small but still
significant amount, it is reasonable to operate with $\delta a = 1-a$. 
% Certain quantities
% important for accretion onto a black hole depend on different powers 
% of $\delta a$ when rotation parameter is close to unity.

A relatively strong limit upon the rotational parameter is set by selective
capture of the radiation of the disc
pointed out by \citet{thorne}. For the thin radiatively efficient disc model
\citep{SS73,NT73}, black hole spin-up proceeds towards the equilibrium value of
$a\simeq 0.998$ (or $\delta a \simeq 2\times 10^{-3}$).

While for radiatively efficient thin disc accretion is relatively 
well understood, 
there is still lack in understanding of geometrically thick
discs and accretion flows that are generally radiatively inefficient and hence
may spin up black holes to higher values of $a$ \citep{abram80, sadowski11}. The
non-Keplerian nature of thick radiatively inefficient 
flows may revert the effect by
lowering the specific angular momentum. Equilibrium spin values
found by \citet{PGII} are considerably lower ($a_{eq}\sim 0.8\div 0.9$). 
 Inclusion of different effects such as magnetic stresses and minor
  mergers (in spin evolution of massive black holes) 
  also leads to the relatively small values of $a_{eq}\sim 0.9$
  \citep{gammie04}. Intermediate Kerr parameters are also generally found in
  fitting the observational data \citep{li05}. 
Here, I will consider primarily the effect of radiation capture paying little
attention to 
deviations from Keplerian law (they are considered approximately in
section~\ref{sec:asympta}) and not considering the effect 
of additional momentum transfer (for instance, by magnetic fields). 

%In reality radiatively efficient and inefficient accretion flows may coexist
%for one object, primarily in the form of a nearly standard outer disc with
%optically thin advection-dominated inner parts (see for example
%\citet{meyers00}). 
There are accretion flow models where radiatively efficient and inefficient
parts coexist,  primarily in the form of a nearly standard outer disc with
optically thin advection-dominated inner parts (see for example
\citet{meyers00}). 
In the inner parts of thin accretion discs in X-ray binaries, radiation pressure
dominates over gas pressure, and the transition from the standard disc to an
optically thin flow should be accompanied by more or less abrupt emission of
the internal energy stored in radiation trapped inside the disc. 
{ More gradual transition to an optically-thin flow should be still
  accompanied by radiation of all the trapped emission, but the shape of the
  photosphere will be more complex. Below, I will consider the inner disc face
  consisting either of one cylindrical surface of a constant radius or of two
  conical surfaces inclined by some angle $\eta$ toward the surface of the
  thin disc. The effect of the transition photosphere }
becomes more pronounced with growing mass accretion rate and is enhanced by
radial advection of trapped radiation. The radial photosphere position
is determined
either by disc evaporation \citep{honma} or by the sonic surface situated
close to $r_{ISCO}$ \citep{penna}. In the latter case, the disc becomes
transparent
because of the density drop after transition to free-fall regime.
Since I restrict myself to the standard disc model with Keplerian rotation law, 
I will not consider solutions with the transition radius situated
inside the last stable orbit. 

\citet{honma} estimates the transition radius due to evaporation as:

$$
R_{tr} \simeq 2.1\times 10^3 \frac{\alpha^4}{\mdot^2} \frac{GM}{c^2}
$$

Where $\mdot = \dot{M}c^2/L_{Edd}$, therefore the normalization in this
expression is different from that used in the original work. Strong dependence
on the poorly known viscosity parameter $\alpha$  makes a broad range of
truncation radii possible. Transition radius will be treated as a free
parameter spanning a broad range of values between the last stable orbit and
several tens of $GM/c^2$. 

The primary goal of this work is to estimate how does the maximal
possible spin depend on the properties of the inner disc such as disc
truncation and existence of a radial photosphere. 
In the next section I describe the technique used to calculate the
radiation braking term based on the method used by \citet{thorne}. In section
\ref{sec:res}, I report the results obtained for the general case of an
accretion disc truncated from inside and the case
of a geometrically and optically thick disc where only photons emitted from its
inner rim may reach the black hole. 
In section \ref{sec:disc}, I discuss the implications and
limitations of my results.

\section{Calculation technique}\label{sec:tech}

\subsection{Local disc radiation field}\label{sec:rfield}

Let us consider that in the co-moving frame, a unit surface element of disc
surface radiates some known energy flux $F=F(R, a)$. For instance, for the
standard thin accretion disc this flux equals:

\begin{equation}\label{E:fsd}
F_{SD} = \frac{3}{8\pi} \frac{GM\dot{M}}{R^3}
\frac{\mathcal{Q}}{\mathcal{B}\sqrt{\mathcal{C}}} = \frac{3}{2}
\frac{c^5}{\varkappa GM}
\frac{\mdot}{r^3} \frac{\mathcal{Q}}{\mathcal{B}\sqrt{\mathcal{C}}}
\end{equation}

Below I will use dimensionless variables $r = R c^2 / GM$, $\mdot =
\dot{M}c^2/L_{Edd}$, where $L_{Edd}$ is Eddington luminosity, $\varkappa$ is
(Thomson) opacity.  
The calligraphic letters denote the coefficients used in the relativistic thin
disc model as given by \citet{penna}.
%, see Appendix~\ref{sec:app}. 
% If accretion is radiatively inefficient, the flux is smaller. 
This flux may be converted to the
fluxes of energy-at-infinity and angular momentum only if the angular
dependence of the intensity of
the outcoming radiation is known. This distribution is different, for
instance, for
optically-thin and optically thick discs and for discs with
contributions of different opacity sources  in the atmosphere. An interesting
possibility is the possible inclination of the disc photosphere caused by
disc thickness dependence on radius. 

Locally measured flux leaving the disc in a unit
solid angle may be expressed as $F \times i(\Theta,\Phi)$, 
where $\int i d\Omega =1$ and $\Theta$ and $\Phi$ characterize the
direction in the frame co-rotating with the disc, $d\Omega = \sin \Theta
d\Theta d\Phi$ (see details in Appendix~\ref{sec:app}). This normalization is
different from the intensity normalization used by \citet{thorne} by a factor
of $\cos \Theta$. 
For an isotropic source, $i=1/4\pi$.
 For the more general case of radiation field symmetric with respect to some
 axis  $i=i(\mu)$, where $\mu = \cos\eta \cos \Theta - \sin \eta \sin \Theta
 \cos \Phi$  for the axis lying in the plane perpendicular to the direction of
 disc rotation and inclined by some angle $\eta$ with respect to the vertical.
% For the more general case of radiation
%field symmetric with respect to some axis lying in the plane perpendicular to
%the direction of disc rotation and inclined by some angle $\eta$ with
%respect to the vertical, $i=i(\mu)$, where $\mu = \cos\eta \cos \Theta -
%\sin \eta \sin \Theta \cos \Phi$. 
Minus sign means that we consider $\eta$
positive if the face of the inclined photosphere is oriented toward the black
hole. In
particular, $\mu = \cos\Theta$ if the disc surface is horizontal and $\mu = -
\sin \Theta \cos \Phi$ for the inner disc face case. For a thin static
plane-parallel photosphere:

$$
i(\Theta, \Phi) =
 \left\{ 
\begin{array}{lc}
\displaystyle
\frac{\mu}{2\pi} & \mbox{ \footnotesize Lambert's cosine law}\\
\\
\displaystyle
\frac{3}{7\pi}\mu\times \left(1+2\mu\right)  
& \mbox{ \footnotesize atmosphere affected by electron-scattering opacity}\\
\end{array}
\right.
$$

The only place where the $I \propto 1+2\mu$ law for outcoming
intensity is derived seems to be the Chandrasekhar's monograph on radiation
transfer \citep{electrochandra}. For $\mu<0$, I assume $i=0$ in both cases. 

Outer parts of accretion discs may have considerable thickness due to disc
thickness dependence on radius and due to irradiation effects (flaring discs,
see \citet{SS73}), but their impact on the black hole is smaller by a
factor of $1/r$. 
In the inner parts of the disc, strong but poorly known dependence of disc
thickness on radius creates inclined portions of disc photosphere. 
% Their inclination angles based on the thin disc theory 
% will be estimated in appendix \ref{sec:NTincl}.
Here, I assume that disc truncation is abrupt enough to make cylindrical
photosphere with $\eta =0$.

\begin{figure*}
 \centering
\includegraphics[width=0.9\textwidth]{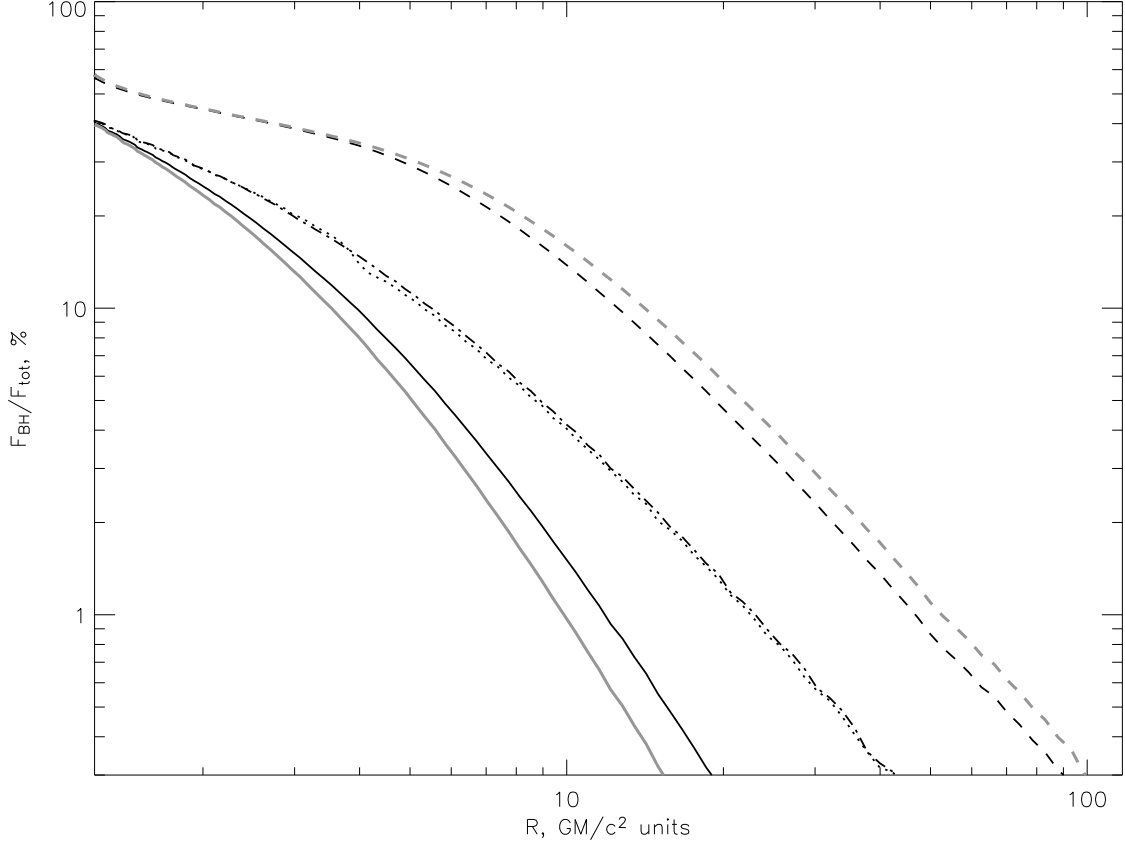}
\caption{ Co-moving flux fraction absorbed by the black hole as a function of
  distance. Solid lines correspond to the thin disc case, 
  dotted for isotropic emission { (optically thin case)}, 
dashed for radial photosphere {(transition region)}. 
  Hereafter grey lines correspond to the case of electron-scattering
  atmosphere solution. { The dot-dashed curve corresponds to the optically
    thin case without the effects of disc thickness. Kerr parameter is set to
    $a=0.998$. } 
%Thick grey lines correspond to intensity distribution
%  affected by electron scattering. 
}
\label{fig:adsd}
\end{figure*}

\subsection{Radiation capture braking term}\label{sec:rbrake}

To calculate the effect of captured radiation upon black hole evolution, it is
convenient to perform integration over solid angles in the co-moving
frame and then over the disc surface. The integrand is $\sqrt{-g} F n_t$ in
the case of energy and  $\sqrt{-g} F n_\varphi$ in the case of angular
momentum, where $n_\varphi$ and $n_t$ are the components of the momentum of a
photon having unit energy in the orbiting frame (see Appendix~\ref{sec:app}).
The multiplier $\sqrt{-g} = \alpha \times
\sqrt{g_{\varphi\varphi} g_{zz} g_{rr}}=r $  takes into account the difference
in the proper and coordinate time ($\alpha$) and for the elementary disc
surface area. First integration should
be performed only over the photon trajectories that finally encounter the
black hole. Condition for hitting the black hole was considered by
\citet{thorne} in terms of effective potential. 
{ Since some parts of the disc may be geometrically thick, photons will be
  at average emitted at some distance from the equatorial plane of the
  disc, equal to the disc height for the standard disc and smaller
  for the optically thin and transition cases. This may be important for the
  condition of
  hitting the black hole that is sensitive to the Carter constant value. All
  the other effects are $O(H/R)^2$ or smaller. Carter's constant is  
  larger and may be evaluated as: 

$$
Q^2 = r^2 n_{z}^2 + \mu_0^2 \times \left(
\frac{n_\varphi^2}{1-\mu_0^2}-a^2\right),
$$

where $n_i$ are normalized coordinate components of the photon momentum (see
Appendix~\ref{sec:app}), $\mu_0$ is calculated as $\mu_0 = H/R$, where 
$H=H_{SD} = \frac{3}{2}\mdot \frac{GM}{c^2}
\times \mathcal{Q} \mathcal{R}^{-1} \mathcal{B}^{-1} \mathcal{C}^{-1/2}$ for
the thin disc case,
$H_{SD}\times \sqrt{2/5}$ for the transition region (see below this section) 
and $H = R/4$ for the optically thick
case. This allows to approximately evaluate the effects of disc thickness. } 

In figure \ref{fig:adsd}, I show the relative fraction of intensity absorbed
by the BH as a function of dimensionless radius.  Optically thin disc is
evidently much more efficient due to different angular dependence of intensity
and besides this has different scaling for large radii. 
At some large dimensionless radius $r$, the directions intercepted by the 
black hole cover a
solid angle of $\Omega_{BH} \sim \pi r^{-2}$ near $\cos \Theta =0$. 
%Hereafter
%lowercase $r$ will be used for normalized radii, $r=Rc^2/GM$.
% and is
%shifted due to Doppler effect by $\sim r^{-1/2}$ in the direction of disc
%rotation. 
Flux fraction absorbed by the black hole scales approximately $\sim \langle
|\cos \Theta|\rangle \times \Omega_{BH} \propto r^{-3}$ in the thin disc case and
$\sim \Omega_{BH} \propto r^{-2}$ in the optically thin case and for a
radially-oriented photosphere. In the last case the gain is even higher
because the radiation is channelled toward the black hole. 
{
In figure ~\ref{fig:adsd}, the dotted and dot-dashed curves correspond to the
optically thin disc case for $H/R =0$ and $H/R=0.5$. 
The amplitude of disc thickness effects 
remains of the order of $(H/2R)^2 \simeq 6\%$ for the considered range of radial
distances. 
}

Radiation contributions to black hole mass growth and angular momentum
evolution are:

$$
\left(\frac{dM}{dt}\right)_{rad} = \frac{1}{c^2} \int_{\mbox{\small disc surface}} F(R) \left( \int_{\mbox{\small co-moving
  }\Omega} i(\Theta, \Phi, R) n_t C_{BH}(\Theta, \Phi, R) d\Omega \right) RdRd\varphi
$$

$$
\left(\frac{dJ}{dt}\right)_{rad} = \frac{GM}{c} \int_{\mbox{\small disc
    surface}} F(R) \left( \int_{\mbox{\small co-moving
  }\Omega} i(\Theta, \Phi, R) n_\varphi C_{BH}(\Theta, \Phi, R) d\Omega
\right) RdRd\varphi
$$

Here, $dA = RdRd\varphi$ is surface area element, $C_{BH}=1$ if the photon
encounters the black hole and $0$ otherwise. The form of $i(\Theta, \Phi, R)$
is set explicitly in accordance with the considered emission regime. Apart from
the case of optically-thin disc and the standard disc case considered
by \citet{thorne}, the most expected picture is a
standard disc truncated from inside with its inner parts replaced by an
optically-thin, geometrically thick flow. In this case, the two above 
integrals may be expressed as sums of three terms corresponding to the inner
transparent part, the inner face of the disc and the outer standard
thin disc. I treat the radial coordinate of the inner face of the disc is
a free
parameter and assume that it coincides with the transition radius $R_{tr}$ 
dividing the
optically thin and optically thick parts of the disc.

The two quantities conserved along the path of the photon, $-n_t$ and
$n_\varphi$, are calculated as functions of its comoving-frame properties (see
Appendix~\ref{sec:app}):

\begin{equation}\label{E:nt}
- n_t = \frac{1}{\sqrt{\mathcal{C}}} \times \left(
\mathcal{G} + \sqrt{\frac{\mathcal{D}}{r}} \sin \Theta \sin \Phi \right) 
\end{equation}

\begin{equation}\label{E:nphi}
n_\varphi =  
\frac{1}{\sqrt{\mathcal{C}}} \times \left(
\sqrt{r}\mathcal{F} + r\mathcal{B} \sqrt{\mathcal{C}} 
\sin \Theta \sin \Phi \right) 
\end{equation}

For the optically thin part, I will assume the accretion disc flux
to be equal to the standard disc flux $F_{SD}$ multiplied by a constant 
factor of $\epsilon < 1 $. This is presumably a very much
simplified picture but sufficient to estimate the principal effect of disc
truncation in its inner parts.

Kerr parameter evolution is governed by a first-order equation of the form:

$$
\frac{da}{dt} = \frac{c}{GM^2}\frac{dJ}{dt}- \frac{2a}{M} \frac{dM}{dt} 
$$

To exclude the black hole mass from the right-hand side it is convenient to
divide the expression by $\frac{1}{M}\frac{dM}{dt}$. 
This leads to the following expression:

\begin{equation}\label{E:spinup}
\frac{da}{d\ln M} =
\frac{c^3}{GM} \frac{L^\dagger+\left(dJ/dt\right)_{rad}/\dot{M}}{E^\dagger+\left(dE/dt\right)_{rad}/\dot{M}} - 2a 
\end{equation}

$L^\dagger$ and $E^\dagger$ are the net angular momentum and
energy-at-infinity at the last stable orbit. Note that $\dot{M}$ is rest-mass
accretion rate and is not equal to $dM/dt$.
The two terms produced by the absorbed radiation may be written as follows:

\begin{equation}\label{E:dedt:gen}
\displaystyle
\begin{array}{l}
\left(\displaystyle \frac{dE}{dt}\right)_{rad} = \int_{R_{in}}^{R_{tr}} R F(R) dR 
\int_{\Omega} n_t C_{BH}(\Theta, \Phi, R) d\Omega + \\
\qquad{}\qquad{}+L_{tr}
\times \int_{\Omega} i_r(\Theta, \Phi) n_t C_{BH}(\Theta, \Phi, R_{tr})
d\Omega-L_{diff}\times \int_{\Omega} n_\varphi C_{BH}(\Theta,
\Phi, R_{tr}+\Delta R) i_s(\Theta,\Phi) d\Omega+\\
\qquad{}\qquad{}\qquad{}+ \int_{R_{tr}}^{+\infty} R F(R) dR 
\int_{\Omega} n_t C_{BH}(\Theta, \Phi, R) i_s(\Theta,\Phi) d\Omega\\
\end{array}
\end{equation}

% The two terms produced by the absorbed radiation may be written as follows:

\begin{equation}\label{E:djdt:gen}
\displaystyle
\begin{array}{l}
 \left(\displaystyle \frac{dJ}{dt}\right)_{rad} \times {\displaystyle
   \frac{c}{GM} } =  \int_{R_{in}}^{R_{tr}} R F(R) dR 
\int_{\Omega} n_\varphi C_{BH}(\Theta, \Phi, R) d\Omega + \\
\qquad{}\qquad{} +L_{tr}
\times \int_{\Omega} i_r(\Theta, \Phi) n_\varphi C_{BH}(\Theta, \Phi, R_{tr})
d\Omega-L_{diff}\times \int_{\Omega} n_\varphi C_{BH}(\Theta,
\Phi, R_{tr}+\Delta R) i_s(\Theta,\Phi) d\Omega +\\
\qquad{}\qquad{}\qquad{}+ 4\pi \int_{R_{tr}}^{+\infty} R F(R)(R)dR 
\int_{\Omega} n_\varphi C_{BH}(\Theta, \Phi, R) i_s(\Theta,\Phi)
d\Omega\\
\end{array}
\end{equation}

Here, $i_s$ and $i_r$ are the normalized intensities for the outer thin disc
($\eta =0$) and for the radial photosphere ($\eta = \pi/2$),
respectively. Their dependence on the angular variables for different cases
was considered in section \ref{sec:rfield}.
% { Correction multiplier
%  $k_{corr}$ is introduced to account for some part of the radiation
%  (contributing to the second term in the above equations) of the
%  standard disc diffusing out of the inner disc face. Instead of calculating
%  the exact form of $k_{corr}$ or setting some {\it ad hoc} form for it, we  }

Co-rotating luminosity $L_{tr}$ of the inner face of the disc may be expressed as:

\begin{equation}\label{E:ltr}
\begin{array}{l}
L_{tr} = L_{diff} + L_{adv} \simeq 4\pi R H F_{SD} +  2\pi R \times U_{rad}
v^{\hat{r}} =\\ 
\qquad{} =2\pi R H F_{SD} \times \left( 1+ 4K \frac{1}{c} \left(\frac{T_c}{T_{eff}}\right)^4 v^{\hat{r}} \right) = 2\pi R H
F_{SD} \times \left( 1+ 4Kv^{\hat{r}} \times
\left(1+\frac{3}{8}\tau\right) \right),
 \end{array}
\end{equation}

where $H$ is disc half-thickness, $U_{rad}$ is vertically integrated energy
density in the disc, $v^{\hat{r}}$ is the locally-measured radial velocity in
the disc.  
The first term describes the radiation diffusing out of the inner face of
the disc (hence its effective temperature is close to the local effective
temperature of the disc). { Since the radiation diffusing out of the
  transition photosphere can not contribute to the radiation of the outer
  standard disc, its contribution is subtracted from the radiation of the disc
  (the negative term proportional to $L_{diff}$ in the formulae
  (\ref{E:dedt:gen}) and (\ref{E:djdt:gen}) above) setting $\Delta R =
  H(R_{tr})$. The lacking disc radiation has smaller effect upon black hole
  rotation than the inner disc face since the orientation of the
  photosphere is different. }
The second term corresponds to the radiation
energy advected out of the optically thick region. Temperature ratio
$T_c/T_{eff}$ was obtained in \citet{SS73} by considering vertical radiation
diffusion. Second term becomes
important if the accretion rate is high and radiation trapping works efficiently
outside the transition radius. In this case, it is reasonable to connect
$R_{tr}$ with the sound surface near the last stable orbit rather than with
disc evaporation. 
$K$ multiplier in the above equation takes into account the vertical
structure of the disk. 
Vertical structure of a radiation-supported disc is fairly approximated by a
polytropic model with polytropic index $n \simeq 1$ (see for example
\citet{SSZ}). For $n=1$, $p_{rad} \propto
\left(1-(z/H)^2\right)$ and  $K = 2/3$. This value was used in all the
calculations. The particular value of $K$ varies slightly with the
polytropic index reaching $ 16/35 \simeq 0.457$ for the extreme value of
$n=3$ that makes the relevant spin-down term 
about 10$\div$20\% smaller. { Mean value of
  $(z/H)^2$ for $n=1$ is $2/5$ that justifies the choice of $\mu_0=H/R=2/5$ 
for the transition region case (see above this section). 

In all the simulations, we use the viscosity $\alpha$ parameter value of
$\alpha=0.1$. 
}
%In particular, for the adiabatic law $p_{rad} \propto
%\left(1-(z/H)^2\right)^{3}$,
%it is equal to $K = 16/35 \simeq 0.457$. Below I will use this
%value of $K$. 
%\frac{\sqrt{\pi}}{6}\frac{\Gamma(1/4)}{\Gamma(3/4)} \simeq 0.87$. Below we
%will adopt $K = 0.87$. 

\subsection{Asymptotic behaviour for $\delta a \ll 1$}\label{sec:asympta}

Let $J^\prime_{rad}$ and $E^\prime_{rad}$ be the
angular momentum and energy emitted and absorbed by the hole for a
unit accreted rest mass. To my knowledge, there are no convenient
expressions for 
$L^\dagger$ and $E^\dagger$ as functions of $a$, but since the considered
Kerr parameter values are very close to unity it is reasonable to apply series
expansion in $\delta a^{1/3}$. To the second order in $\delta a^{1/3}$, using
expressions (\ref{E:ISCO}), (\ref{E:ldag0}) and (\ref{E:edag0}), one obtains:

\begin{equation}\label{E:ldag}
L^\dagger \times \frac{c}{GM} = \frac{2}{\sqrt{3}}\times \left( 1+2^{2/3} \delta a^{1/3} + 2^{-5/6} \delta
a^{2/3}\right)+O(\delta a)
\end{equation}

\begin{equation}\label{E:edag}
E^\dagger \times c^{-2} = \frac{1}{\sqrt{3}}\times \left( 1+2^{2/3} \delta a^{1/3} - \frac{5}{3}
2^{-8/3} \delta a^{2/3}\right)+O(\delta a)
\end{equation}

For Kerr parameter evolution, one obtains through direct substitution of the
above expansions into the spin-up law~(\ref{E:spinup}):

\begin{equation}\label{E:linear}
\frac{da}{d\ln M} \simeq  2 \left[ \left(2^{-5/6} + \frac{5}{3} 2^{-8/3} \right)
\delta a^{2/3} + \sqrt{3} \left( \frac{c}{2GM} J^\prime_{rad} -  \frac{1}{c^2}E^\prime_{rad}\right) \right],
\end{equation}

This expression provides a general scaling for the equilibrium rotation
parameter provided that $\delta a_{eq}$ is small and the impact of radiation
is much smaller than that of accreted matter:

$$
\delta a_{eq} \propto \left( \frac{c}{2GM} J^\prime_{rad} - \frac{1}{c^2} E^\prime_{rad}\right)^{3/2}
$$

Here, $J^\prime_{rad}$ and $E^\prime_{rad}$ scale with the local radiative 
efficiency
of accretion $\epsilon$, that implies $\delta a \propto \epsilon^{3/2}$. For the standard disc case, they do not depend on the mass
accretion rate, but dependence on accretion rate may arise for the inner disc
face. 
%\bigskip
%
%If the black hole is situated far enough, its visible 
%position from the point of view of an observer in the disc will be shifted
%forwards by the angle of $\simeq v_K/c \simeq r^{-1/2}$ 
%due to light aberration. Photons reaching the black hole should be inclined
%backwards by the identical angle with respect to radial direction. 
%General
%relativity effects affecting the black hole image shape and position are
%smaller. The mean photon motion integrals are $-n_t \simeq 1$
%and $n_\varphi = O(r^{-1/2})$, i. e., radiation
%increases the mass of the black hole without changing the
%overall angular momentum that leads to spin-down in terms of Kerr parameter. 
%The photons
%reaching the black hole have zero angular momentum up to the zeroth order in
%$r^{1/2}$.
% Hence we may assume that only the energy term is important for
%black hole spin-down in this case and:
%
%$$
%E^\prime_{rad} \simeq \frac{4\pi}{\dot{M}} \int_R \Omega_{BH}(R) F(R) R dR 
%$$
%
%where $\Omega_{BH}\simeq \pi r^{-2}$ is the solid angle subtended by the black hole. 
While in the thin-disc limit, $E^\prime_{rad}$ and $J^\prime_{rad}$
do not depend on $\mdot$, existence of an inner face with $H
\propto \mdot$ leads
to $J^\prime_{rad} \propto E^\prime_{rad} \propto \mdot$ and 
hence $\delta a_{eq} \propto
\mdot^{3/2}$. Broader applicability of this scaling is supported by the more
comprehensive numerical results given in section \ref{sec:res:edge}.

{

If the disc is non-Keplerian but its inner rim is fixed to the ISCO radius,
deviations from the Keplerian may play the
main role. Spin evolution is then determined (in the $\delta a \ll 1$ limit)
by the following expression:

$$
\frac{da}{d\ln M} \simeq  2 C_1 \left[
  3\times 2^{-2/3} \delta a^{2/3} -  \left(\frac{1}{C_1} -1\right)  \right],
$$

where $C_1$ is the angular momentum at the ISCO in the units of Keplerian
angular momentum. 
Equilibrium spin value may be supported without radiation capture in this
case:

$$
\delta a_{eq} \simeq 2 \times 3^{-3/2} \left( \frac{1}{C_1}-1 \right)^{3/2}
$$
}

\section{Results}\label{sec:res}

\subsection{Inner edge of a thick disc}\label{sec:res:edge}

Most of the internal energy in the inner parts of X-ray binary discs is stored
in the pressure of radiation diffusing upward towards the disc
surface. If the disc abruptly becomes optically thick (that is expected in the
case of disc evaporation or near the ISCO), this radiation escapes due to
diffusion and advection. The luminosity created by this ``transition''
radiation source is estimated by the expression (\ref{E:ltr}) above. This
radiation is much more efficient in spinning down the black hole than standard
disc radiation. 

\begin{figure*}
 \centering
\includegraphics[width=\textwidth]{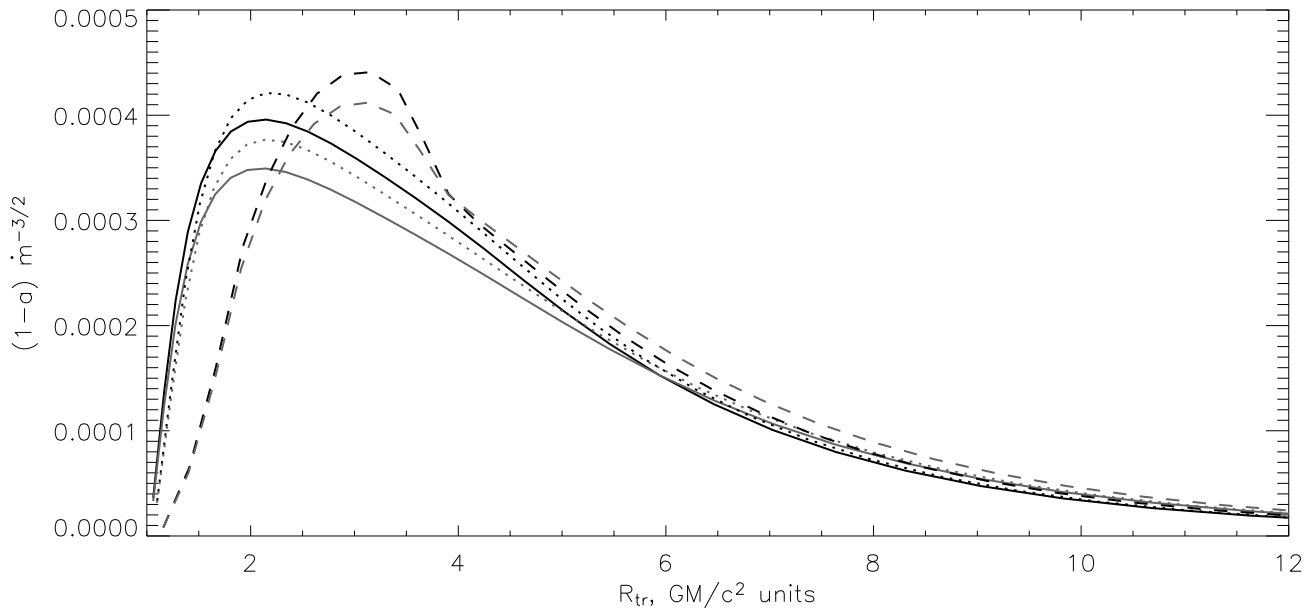}
\caption{ Equilibrium Kerr parameter values (for the ``radial photosphere''
  case considered in section~\ref{sec:res:edge}) normalized by dimensionless
  mass accretion rate $\mdot^{3/2}$ for $\mdot = 0.1$ (solid lines), $1$
  (dotted) and $10$ (dashed).
% Grey lines correspond to the photosphere with
%  the intensity distribution affected by electron scattering. 
}
\label{fig:front}
\end{figure*}

Since disc thickness is proportional to mass accretion rate, the
equilibrium Kerr parameter becomes dependent on the mass accretion
rate. Maximal spin-down occurs for $r\sim 2\div 5$ and results in $\delta
a_{eq} \sim 4\times 10^{-4}\mdot^{3/2}$ (see figure \ref{fig:front}). The
dependence on mass accretion rate is easily explained in the large-radius
limit but holds to $\sim 20\%$ accuracy even if the inner disc rim is close to
the ISCO. One should expect near-critical and mildly super-critical accretion
to be efficient in spinning down the BH to the probable $a_{eq} \sim 0.995$. 

\begin{figure*}
 \centering
\includegraphics[width=\textwidth]{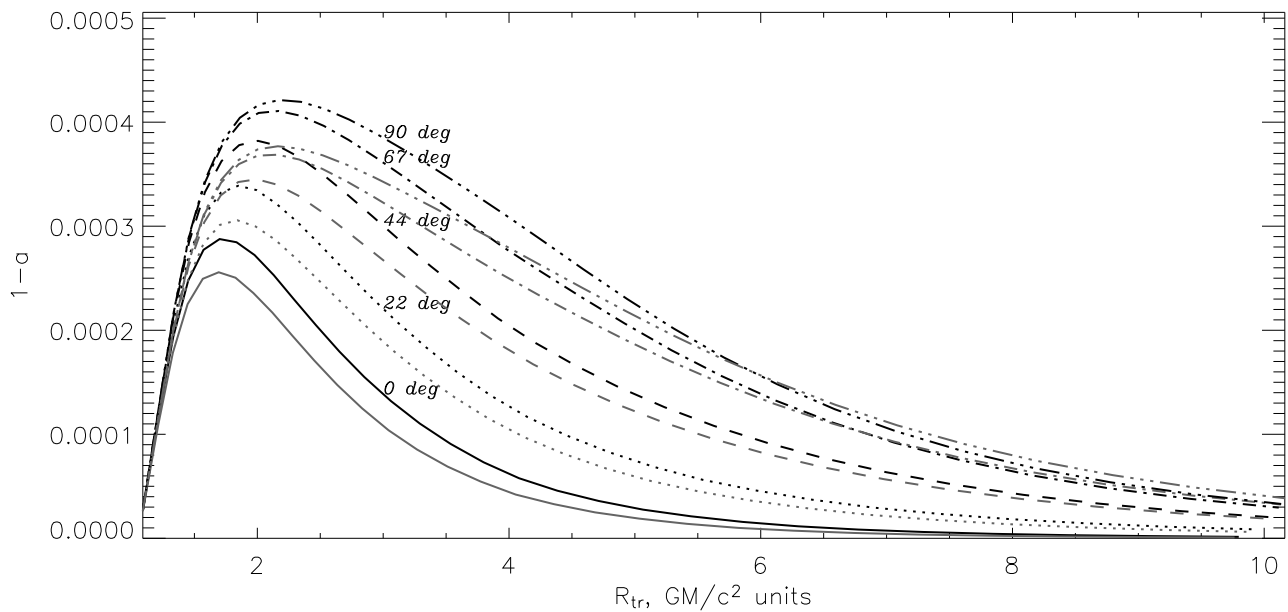}
\caption{ Equilibrium Kerr parameter values for the case of thick disc
  photosphere inclined by some angle $\eta$. Solid, dotted, dot-dashed and
  multiple dot-dashed lines correspond to $\eta = 0, 22, 44, 67$ and $90$
  degrees. Mass accretion rate is fixed to $\mdot =1 $.
}
\label{fig:incl}
\end{figure*}

Since the disc has some non-trivial vertical structure, its inner face is
expected to be not exactly cylindrical. To account for this, I considered a
photosphere inclined by different angles $\eta$ (see figure \ref{fig:incl}; {
 $\eta$ is the angle between the normal to the emitting surface and vertical
direction). The photosphere is assumed symmetric with respect to the
disc plane that implies bi-conical shape of its surface. }

It should be noted that for large spin parameter values, the shape of the
inner parts of the thin relativistic disc deviates strongly from
plane-parallel approximation. It may be shown that
for $\delta a \lesssim 10^{-2}$, the inner parts of the thin disc are inclined
by $\gtrsim 20^\circ$ to the equatorial plane. 

Results of this subsection may be used to make a rough estimate for the
maximal possible $\delta a$ in the optically and geometrically thick
supercritical disc if it is due to some reason (such as high viscosity or
magnetic pressure) truncated outside the ISCO. The
maximal possible disc thickness is $H\sim R$ that corresponds to $\mdot \simeq
2/3\eta(a) \simeq 3$, where $\eta(a)\sim 0.32$ is accretion disc efficiency. 
For $r_{tr} = 2$, this implies $\delta a \simeq 4\times 10^{-4} \times
(3)^{3/2} \sim 2\times 10^{-3}$. Hence, equilibrium
Kerr parameter is
unlikely to become smaller than $\sim 0.998$ through radiation capture from an
optically and geometrically thick disc if the outer disc parts are invisible. 

\subsection{Truncated disc case}

\begin{figure*}
 \centering
\includegraphics[width=\textwidth]{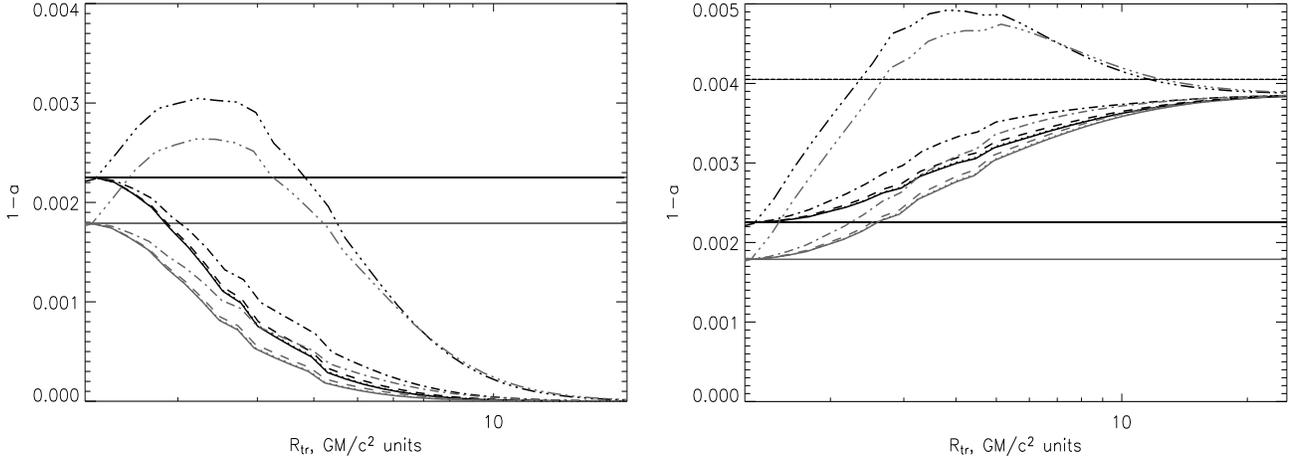}
\caption{ Equilibrium Kerr parameter values as functions of truncation radius
  $r_{tr}$ for different mass accretion rate values ($\mdot = 10^{-2}$,
  $4\times 10^{-2}$, $0.17$, $0.72$ and $3$ are shown by solid, dotted,
  dashed, dot-dashed and double dot-dashed lines, respectively). 
%and different emission
%  regimes (black and grey lines correspond to Lambert and electron-scattering
%  photospheres, respectively). 
  In the left panel, the inner ADAF disk part emits nothing, in
  the right panel its locally emitted flux equals that for the standard
  disc. { Horizontal lines mark the Thorne limits of $\delta a \simeq
   $ 0.0018 and 0.0022 and the maximal possible $\delta a\simeq 0.004$ for the
    optically
    thin case (not taking into account disc thickness, hence this value is
    higher than the limiting $\delta a$ for a hybrid disc).}
}
\label{fig:atr2}
\end{figure*}

In figure~\ref{fig:atr2}, I show the dependence of equilibrium $a$ on the
inner truncation radius $r_{tr}$ in two extreme cases: if the optically thin
disc part emits nothing (in this case, there is no Thorne spin-down term in
the limit $r_{tr} \to \infty$) and if it produces exactly the same amount of
radiation as standard disc, but the radiation is emitted isotropically. An
optically thin disc has a potential for 
somewhat stronger radiative spin-down than the
standard disc, leading {  (if it happens to be radiatively efficient and
  geometrically thin) to}
the equilibrium Kerr parameter value of $\delta a_{eq} \simeq
(4.02\pm 0.05)\times 10^{-3} \epsilon^{3/2}$. 

There is a range of $r_{tr} \sim 1.5\div 5$ where the efficiency of truncated
disc radiation in spinning the black hole down may be 
higher than for the standard disc case. 
The prominent bump at these $r_{tr}$ is produced by the radial
photosphere
(see previous subsection). Its contribution grows rapidly with mass accretion
rate and is responsible for the difference between the individual
curves in figure \ref{fig:atr2}. For mass accretion rates $\mdot \lesssim
0.01$, the variations in spin-down term with $\mdot$ are $\lesssim 20\%$.

The maximal possible $\delta a$ and the position of the maximum
depend on the efficiency of the optically thin part of the flow and change
from $r_{tr} \simeq 3$, $\delta a_{max} \simeq 3\times 10^{-3}$ for
$\epsilon=0$ to $r_{tr} \simeq 4$, $\delta a_{max} \simeq 5\times 10^{-3}$
for $\epsilon=1$. One additional assumption needed for this estimates to work
is direct visibility of the emitting surface of the disc from the black
hole. If the disc launches wind or its inner parts are geometrically 
thicker the efficiency
of radiation-capture mechanism is always smaller than of in the standard disc
case (see previous subsection). 

%\subsection{Compact discs}
%
%In some cases such as accretion from stellar wind, there are reasons to
%consider accretion discs truncated from outside. 
%In this case, the only source
%of radiation is the disc, optical thick of thin, and its impact is always
%lower than that of the disc without the outer boundary (see~\ref{fig:inner}). 
%Most of the spin-down effect is created by the radiation emitted by the inner
%$\sim 3\,GM/c^2$ of the disc. 
%
%\begin{figure*}
% \centering
%\includegraphics[width=0.75\textwidth]{atrshow_inner}
%\caption{ Equilibrium Kerr parameter values for a disc truncated from outside
%  at $r_{tr}$. Solid lines correspond to standard disc, dotted to the
%  optically thin case.  Horizontal lines mark the limits obtained by
%  \citet{thorne} (solid lines) and our limit for an optically thin disc
%  (dotted). 
%}
%\label{fig:inner}
%\end{figure*}
%
%Wind-fed accretion is characteristic for massive X-ray binaries with
%short-living donors. Such a stage is unlikely for the hole to increase its
%mass by considerable amount since evolutionary stage of the donor is at least
%an order of magnitude smaller than the Eddington time of the black hole. Small
%inner discs may appear around isolated black holes accreting from the
%interstellar medium with small velocity and density gradients.  

\section{Discussion}\label{sec:disc}

The effect of radiation spin-down is
probably of little importance because magnetic fields are expected to spin
black holes down much more efficiently \citep{BZ77,uzdensky,gammie04}. 
The role of mechanisms involving magnetic fields depends on the geometry of
the field and on the existence of some extended load (such as jet). Hence the
situation when magnetic fields are insufficient to stop the spin-up should not be
excluded. Unlike the effect of radiation capture, the impact of
Blandford--Znajek and similar process changes smoothly 
when $\delta a$ approaches
zero and may be safely ignored unless it already provides an equilibrium $a<1$
by itself. 

Radiation of optically thin ADAF discs is of little importance for black
hole evolution since the radiation is emitted inefficiently and hence the
amplitude of the radiation contribution is several orders smaller than for the
radiatively-efficient standard disc case.
Potentially interesting case is hyperaccretion through a neutrino-emitting
accretion disc \citep{hyperborodov}. 
For stellar-mass black holes and $\dot{M} \sim
10^{-3}\div 10^{-1}\Msunyr$, accretion disc emits neutrinos in an optically-thin
but radiatively efficient regime. 

Another complication that should be taken
into account in more comprehensive models is deviation from the Keplerian law
possibly important for radiatively inefficient discs. Substantially
sub-Keplerian flows are unable to spin the black hole up to Kerr parameters
where radiation capture effect becomes important \citep{PGII}.
It can be checked that if the net angular momentum at the last stable orbit
differs from Keplerian by a factor of $c<1$, spin-up proceeds up to some
equilibrium value of $a<1$. Estimates made in section~\ref{sec:asympta}
suggest scaling $\delta a_{eq} \simeq 2\times 3^{-3/2} \times (1/c-1)^{3/2}$
in this case. 
Deviations from Keplerian rotation become important if deviations from
Keplerian law are $\gtrsim (1\div 2)\%$. For thick discs with $H/R \gtrsim
0.1$, sub-Keplerian rotation may be a more important factor than radiative
spin-down. Disc rotation faster than Keplerian by several
percent makes it impossible to balance black hole rotation by radiation
capture. Super-Keplerian slim discs considered by \citet{sadowski11} 
provide the black hole with matter
having not only higher net angular momentum but also
exceedingly high net energy hence the overall spin-up may be still stopped by
radiation. Inner structure for these accretion disc models is profoundly
different from the thin-disc approximation and also shows strong dependence on
viscosity. 

%For an optically thin disc with a constant fraction $\epsilon$ 
%of radiated energy, the
%equilibrium value of $a$ is $a_{eq} \simeq 1-(1-a_0) \times \epsilon^{3/2}$ 
%where $a_0\simeq 0.996$ is somewhat smaller than the values expected for
%standard discs. The scaling with $\epsilon$ here works whenever $\delta a \ll
%1$, $E^\prime_{rad} \ll E^\dagger$ and $L^\prime_{rad}\ll L^\dagger$. 

\section{Conclusions}

I come to the conclusion that radiative spin-down is sensitive to
the geometry and optical depth of the emitting material. Existence of a
radially oriented disc photosphere at several gravitational radii may increase
the spin-down term by about a factor of 1.5 for large (near-critical) accretion
rates in the disc if the inner edge of the standard disc part lies in the
range $(2\div 4) GM/c^2$. In other cases the effect of disc radiation  is much
smaller due to lower radiative efficiency. { Non-Keplerian rotation becomes
  more important than radiation capture if deviations from Keplerian law
  exceed 1$\div$2\%. }

{
\section*{Acknowledgments}

This work was supported by the RFBR grant 12-02-00186-а. Author thanks
the anonymous referee for valuable comments. 
}

\appendix
% \twocolumn

\section{Some essential properties of the thin relativistic disc model and
  transition from the orbiting frame}\label{sec:app}

This Appendix was introduced only for reference and does not contain any
original research results. Since all the lengths scale either with the black
hole mass or with the Kerr parameter, I  assume here $GM = c = 1$. Kerr metric
in the Boyer-Lindquist
coordinates may be expressed near the equatorial plane as:

$$
ds^2 = -\alpha^2 dt^2 + \frac{\Sigma^2}{r^2}\left(d\varphi-\omega_{LT}
dt\right)^2 + \frac{r^2}{\Delta} dr^2+dz^2
$$

Here:

$$
\alpha^2 = \frac{r^2 \Delta}{\Sigma^2}
$$

$$
\Sigma^2 = r^4+r^2a^2 + 2 r a^2 = r^4 \mathcal{A}
$$

$$
\Delta = r^2-2r+a^2
$$

Lense-Thirring precession frequency:

$$
\omega_{LT} = \frac{1}{r^{3/2}+a} = \mathcal{B}^{-1} r^{-3/2} 
$$

The theory of relativistic thin accretion disk as introduced by
\citet{NT73,PT74} operates a series of auxiliary factors depending on the
radial coordinate and rotation parameters $a$ that I here denote with
calligraphic letters following the notation given by \citet{penna}.
% Instead of the normalized radial coordinate $r$,
%sometimes we will use $x=\sqrt{r}$.
%
%$$
%\mathcal{A} =  1+\frac{a^2}{r^2} + \frac{2a^2}{r^3}
%$$
%
%$$
%\mathcal{B} =  1+\frac{a^2}{r^{3/2}}
%$$
%
%$$
%\mathcal{C} =  1-\frac{3}{r}+\frac{a^2}{r^2}
%$$
%
%$$
%\mathcal{D} =  1-\frac{2}{r}+\frac{a^2}{r^2}
%$$
%
%$$
%\mathcal{E} = 1+ \frac{4a^2}{r^2}-\frac{4a^2}{r^3}+\frac{3a^4}{r^4}
%$$
%
%$$
%\mathcal{F} = 1- \frac{2a}{r^{3/2}}+\frac{a^2}{r^2}
%$$
%
%$$
%\mathcal{G} =  1-\frac{2}{r}+\frac{a^2}{r^{3/2}}
%$$
%
%$$
%\mathcal{R} = \frac{F^2}{C}-\frac{a^2}{x^2} \left(\frac{G}{\sqrt{C}}-1\right)
%$$
%
%$$
%\mathcal{Q} = \frac{\mathcal{B}}{x\sqrt{\mathcal{C}}} \times \left( x-x_0 -
%\frac{3}{2} a \ln \frac{x}{x_0}-A_1 -A_2 -A_3\right),
%% \mathcal{Q} = \frac{3}{2}\times \frac{x-x_0-\frac{3}{2}a\ln (x/x0) - A_1 -A_2
%%  -A_3}{x-3/x+2a/x^2},
%$$
%
%where $x_0 = \sqrt{r_{ISCO}}$, 
%
%$$
%A_1 = \frac{3(x_1 - a)^2)}{x_1 (x_1-x_2) (x_1-x_3)}\ln \left(\frac{x-x_1}{x_0-x_1} \right)
%$$
%
%and $A_2$ and $A_3$ are obtained as cyclic permutations of $A_1$, $x_1$, $x_2$
%and $x_3$ are solutions of the cubic equation $x^3 -3x+2a=0$. 

\bigskip

Net angular momentum and energy on equatorial Keplerian orbits are expressed
as:

\begin{equation}\label{E:ldag0}
L^\dagger = \frac{\sqrt{r}\mathcal{F}}{\sqrt{\mathcal{C}}}
\end{equation}

\begin{equation}\label{E:edag0}
E^\dagger = \frac{\mathcal{G}}{\sqrt{\mathcal{C}}}
\end{equation}

The inner rim of the disc is set by the innermost stable orbit radius that may
be expressed as follows:

\begin{equation}\label{E:ISCO}
r_{ISCO} = 3+Z_2 - \sqrt{(3-Z_1)\left(3+Z_1+2Z_2\right)},
\end{equation}

where:

$$
Z_1 = 1+\left(1-a^2\right)^{1/3} \left(
\left(1+a\right)^{1/3}+\left(1-a\right)^{1/3}\right)
$$

$$
Z_2 = \sqrt{3a^2+Z_1^2}
$$

If $\delta a \ll 1$, the last stable orbit radius is a smooth function of
$\delta a^{1/3}$ that justifies the approximation I use in section
\ref{sec:asympta}. 

\bigskip

Let us consider the orbiting frame moving with the matter with the
four-velocity of $u^i$, $u^\theta = u^r =0$,  $u^\varphi = \Omega
u^t$. Normalization yields

$$
u^t = \frac{1}{\sqrt{\alpha^2-g_{\varphi\varphi} (\Omega -\omega_{LT} )^2}}
$$

I use the orbiting frame tetrad (see for example \citet{NT73}) that is
more convenient to express in terms of covariant (1-form) basis:

\begin{equation}\label{E:tetrad:t}
%\begin{array}{l}
\vector{\omega}^{\hat{t}} = \frac{1}{u^t} \times\left( dt
-\frac{\alpha^2}{g_{\varphi\varphi}}\left(\Omega - \omega_{LT}\right) \times
\left(d\varphi - \Omega dt\right) \right) = %\\\qquad{} = 
\mathcal{C}^{-1/2} \times 
\left( \mathcal{G} dt - \sqrt{r} \mathcal{F} d\varphi\right) %\\
% \vector{e}_{\hat{t}} = u^t \times \left( \pardir{t}{} + \Omega
% \pardir{\varphi}{}\right)
%\end{array}
\end{equation}

\begin{equation}\label{E:tetrad:phi}
%\vector{e}_{\hat{\varphi}} = u^t \left(
%\frac{\alpha}{\sqrt{g_{\varphi\varphi}}} \pardir{\varphi}{} + \frac{\sqrt{g_{\varphi\varphi}}}{\alpha} (\Omega - \omega_{LT}) \left( \pardir{t}{} + \Omega \pardir{\varphi}{} \right) \right)
%\begin{array}{l}
\vector{\omega}^{\hat{\varphi}} = \frac{1}{u^t}
\times\frac{\sqrt{g_{\varphi\varphi}}}{\alpha} \left( -\Omega dt + d\varphi
\right) = % \\\qquad{} = 
\mathcal{C}^{-1/2} \times 
\left( -\sqrt{\frac{\mathcal{D}}{r}}  dt +
r\mathcal{B}\sqrt{\mathcal{D}}d\varphi\right) %\\\end{array}
\end{equation}

\begin{equation}\label{E:tetrad:r}
\vector{\omega}^{\hat{r}} = \sqrt{g_{rr}} dr = \mathcal{D}^{-1/2} dr
%\vector{e}_{\hat{r}} = \frac{1}{\sqrt{g_{rr}}} \pardir{r}{}
\end{equation}

\begin{equation}\label{E:tetrad:t}
\vector{\omega}^{\hat{z}} = dz  
\end{equation}

In the orbiting frame, a photon is characterized by the unit vector
$n^{\hat{a}}$: 

% [do I use positive metric signature everywhere????]

$$
n_{\hat{t}} = -1  \mbox{; ~~~}  
n_{\hat{\varphi}} = \sin\Theta \sin\Phi  \mbox{; ~~~}  
n_{\hat{r}} = \sin\Theta \cos\Phi  \mbox{; ~~~}  
n_{\hat{z}} = \cos\Theta   \mbox{; ~~~}  
$$

These quantities may be connected to the coordinate-frame
vector components as $n_{\hat{a}} = n_i e_{\hat{a}}^i$. This allows to express
the energy-at-infinity $-u_t$ and angular momentum $u_\varphi$ of a given
photon as:

$$
- n_t = -\omega^{\hat{a}}_t n_{\hat{a}} 
= \frac{1}{\sqrt{\mathcal{C}}} \times \left(
\mathcal{G} + \sqrt{\frac{\mathcal{D}}{r}} \sin \Theta \sin \Phi \right) 
$$

$$
n_\varphi = \omega^{\hat{a}}_\varphi n_{\hat{a}} = 
\frac{1}{\sqrt{\mathcal{C}}} \times \left(
\sqrt{r}\mathcal{F} + r\mathcal{B} \sqrt{\mathcal{D}} 
\sin \Theta \sin \Phi \right) 
$$

These quantities multiplied by the local intensity and integrated over the
solid angle give the energy-at-infinity and angular momentum fluxes (see
section~\ref{sec:rbrake}).

\bibliographystyle{mn2e}
\bibliography{mybib}

\hyphenation{Post-Script Sprin-ger}
\begin{thebibliography}{}

\bibitem[\protect\citeauthoryear{{Abramowicz} \& {Lasota}}{{Abramowicz} \&
  {Lasota}}{1980}]{abram80}
{Abramowicz} M.~A.,  {Lasota} J.~P.,  1980, \actaa, 30, 35

\bibitem[\protect\citeauthoryear{{Bardeen}}{{Bardeen}}{1970}]{bardeen70}
{Bardeen} J.~M.,  1970, \nat, 226, 64

\bibitem[\protect\citeauthoryear{{Bardeen} \& {Petterson}}{{Bardeen} \&
  {Petterson}}{1975}]{BP75}
{Bardeen} J.~M.,  {Petterson} J.~A.,  1975, \apjl, 195, L65

\bibitem[\protect\citeauthoryear{{Blandford} \& {Znajek}}{{Blandford} \&
  {Znajek}}{1977}]{BZ77}
{Blandford} R.~D.,  {Znajek} R.~L.,  1977, \mnras, 179, 433

\bibitem[\protect\citeauthoryear{{Chandrasekhar}}{{Chandrasekhar}}{1960}]{electrochandra}
{Chandrasekhar} S.,  1960, {Radiative transfer}.
New York: Dover, 1960

\bibitem[\protect\citeauthoryear{{Chen} \& {Beloborodov}}{{Chen} \&
  {Beloborodov}}{2007}]{hyperborodov}
{Chen} W.-X.,  {Beloborodov} A.~M.,  2007, \apj, 657, 383

\bibitem[\protect\citeauthoryear{{Gammie}, {Shapiro} \& {McKinney}}{{Gammie}
  et~al.}{2004}]{gammie04}
{Gammie} C.~F.,  {Shapiro} S.~L.,    {McKinney} J.~C.,  2004, \apj, 602, 312

\bibitem[\protect\citeauthoryear{{Honma}}{{Honma}}{1996}]{honma}
{Honma} F.,  1996, \pasj, 48, 77

\bibitem[\protect\citeauthoryear{{Li}, {Zimmerman}, {Narayan} \&
  {McClintock}}{{Li} et~al.}{2005}]{li05}
{Li} L.-X.,  {Zimmerman} E.~R.,  {Narayan} R.,    {McClintock} J.~E.,  2005,
  \apjs, 157, 335

\bibitem[\protect\citeauthoryear{{Meyer}, {Liu} \& {Meyer-Hofmeister}}{{Meyer}
  et~al.}{2000}]{meyers00}
{Meyer} F.,  {Liu} B.~F.,    {Meyer-Hofmeister} E.,  2000, \aap, 361, 175

\bibitem[\protect\citeauthoryear{{Novikov} \& {Thorne}}{{Novikov} \&
  {Thorne}}{1973}]{NT73}
{Novikov} I.~D.,  {Thorne} K.~S.,  1973, in {Dewitt} C.,  {Dewitt} B.~S.,  eds,
  Black Holes (Les Astres Occlus) {Astrophysics of black holes.}.
pp 343--450

\bibitem[\protect\citeauthoryear{{Page} \& {Thorne}}{{Page} \&
  {Thorne}}{1974}]{PT74}
{Page} D.~N.,  {Thorne} K.~S.,  1974, \apj, 191, 499

\bibitem[\protect\citeauthoryear{{Penna}, {S\c{a}dowski} \& {McKinney}}{{Penna}
  et~al.}{2012}]{penna}
{Penna} R.~F.,  {S\c{a}dowski} A.,    {McKinney} J.~C.,  2012, \mnras, 420, 684

\bibitem[\protect\citeauthoryear{{Popham} \& {Gammie}}{{Popham} \&
  {Gammie}}{1998}]{PGII}
{Popham} R.,  {Gammie} C.~F.,  1998, \apj, 504, 419

\bibitem[\protect\citeauthoryear{{S{\c a}dowski}, {Bursa}, {Abramowicz},
  {Klu{\'z}niak}, {Lasota}, {Moderski} \& {Safarzadeh}}{{S{\c a}dowski}
  et~al.}{2011}]{sadowski11}
{S{\c a}dowski} A.,  {Bursa} M.,  {Abramowicz} M.,  {Klu{\'z}niak} W.,
  {Lasota} J.-P.,  {Moderski} R.,    {Safarzadeh} M.,  2011, \aap, 532, A41

\bibitem[\protect\citeauthoryear{{Shakura} \& {Sunyaev}}{{Shakura} \&
  {Sunyaev}}{1973}]{SS73}
{Shakura} N.~I.,  {Sunyaev} R.~A.,  1973, \aap, 24, 337

\bibitem[\protect\citeauthoryear{{Shakura}, {Sunyaev} \&
  {Zilitinkevich}}{{Shakura} et~al.}{1978}]{SSZ}
{Shakura} N.~I.,  {Sunyaev} R.~A.,    {Zilitinkevich} S.~S.,  1978, \aap, 62,
  179

\bibitem[\protect\citeauthoryear{{Thorne}}{{Thorne}}{1974}]{thorne}
{Thorne} K.~S.,  1974, \apj, 191, 507

\bibitem[\protect\citeauthoryear{{Uzdensky}}{{Uzdensky}}{2005}]{uzdensky}
{Uzdensky} D.~A.,  2005, \apj, 620, 889

\end{thebibliography}

\label{lastpage}

\end{document}